\documentclass[12pt,jcp,preprint,superscriptaddress]{revtex4}

\usepackage{amsmath}
\usepackage{amssymb}

\usepackage{graphicx}

\usepackage{color}
\usepackage{algorithmic}
\usepackage{algorithm}

\graphicspath{{./Figures/}}

\def\I{{\bf 1}}
\def\M{{\mathcal M}}

\usepackage[labelfont=bf,labelsep=period,justification=raggedright]{caption}

\makeatletter
\renewcommand{\@biblabel}[1]{\quad#1.}
\makeatother

\begin{document}

\title{Hierarchical Nystr\"{o}m Methods for Constructing Markov State Models for Conformational Dynamics}
\author{Yuan Yao\footnote{To whom correspondence should be addressed: {\tt {yuany@math.pku.edu.cn}} or {\tt{xuhuihuang@ust.hk}}}
}
\affiliation{School of Mathematical Sciences, LMAM-LMEQF-LMPR, Peking University, Beijing 100871, China}
\author{Raymond Z. Cui}
\affiliation{Department of Chemistry, Division of Biomedical Engineering, Center of Systems Biology and Human Health, School of Science and Institute for Advance Study, The Hong Kong University of Science and Technology, Clear Water Bay, Kowloon, Hong Kong}
\author{Gregory R. Bowman}
\affiliation{Department of Molecular and Cell Biology, University of California, Berkeley, CA 94720}
\author{Daniel-Adriano Silva}
\affiliation{Department of Chemistry, Division of Biomedical Engineering, Center of Systems Biology and Human Health, School of Science and Institute for Advance Study, The Hong Kong University of Science and Technology, Clear Water Bay, Kowloon, Hong Kong}
\author{Jian Sun}
\affiliation{Mathematical Sciences Center, Tsinghua University, Beijing 100084, China}
\author{Xuhui Huang$^{\mathrm{*}}$}
\affiliation{Department of Chemistry, Division of Biomedical Engineering, Center of Systems Biology and Human Health, School of Science and Institute for Advance Study, The Hong Kong University of Science and Technology, Clear Water Bay, Kowloon, Hong Kong}

\date{\today}

\begin{abstract}
Markov state models (MSMs) have become a popular approach for investigating the conformational dynamics of proteins and other biomolecules.
MSMs are typically built from numerous molecular dynamics simulations by dividing the sampled configurations into a large number of microstates based on geometric criteria.
The resulting microstate model can then be coarse-grained into a more understandable macro state model by lumping together rapidly mixing microstates into larger, metastable aggregates.
However, finite sampling often results in the creation of many poorly sampled microstates.
During coarse-graining, these states are mistakenly identified as being kinetically important because transitions to/from them appear to be slow.
In this paper we propose a formalism based on an algebraic principle for matrix approximation, \emph{i.e.} the Nystr\"{o}m method, to deal with such poorly sampled microstates.
Our scheme builds a hierarchy of microstates from high to low populations and progressively applies spectral clustering on sets of microstates within each level of the hierarchy. It helps spectral clustering identify metastable aggregates with highly populated microstates rather than being distracted by lowly populated states.
We demonstrate the ability of this algorithm to discover the major metastable states on two model systems, the alanine dipeptide and trpzip2 peptide.
\end{abstract}
\maketitle

\section{Introduction}
Markov state models (MSMs) are a powerful approach to model both the thermodynamics and kinetics of proteins and other biomolecules \cite{Zwanzig_1983, Bowman_methods,Hummer_JCP2008, Chodera_JCP2007,Weber_PCCA,Noe_MSMReview, Levy_PNAS2008,Caflisch_JCTC2012, Huang_PNAS2009,Gunsteren_JCTC2011,Roux_2008,Milestoning}.  In an MSM, one decomposes the conformational space of a molecule into metastable states (i.e. long-lived states that correspond to basins in the free energy landscape that ultimately determines a system's structure and dynamics) and calculates a transition probability matrix whose entries specify the probability of transitioning between each pair of states in some time interval, called the lag time of the model.  Using the transition probability matrix, one can then calculate the relaxation of an ensemble of proteins upon the temperature jump \cite{Zhuang_JPC2011,Cui_CurrPChem2012}, find the highest flux paths between pairs of states \cite{E-Eric,MeScVa09,Noe09}, or perform numerous other analyses \cite{Voelz_CurrOpinStructBiol2011,Noe_MSMReview}.  In recent years, MSMs have been successfully applied to study protein folding \cite{Voelz_CurrOpinStructBiol2011,Voelz_JACS2010,Noe09,Bowman_JACS2011,Izaguirre_2010}, RNA folding \cite{Huang_PNAS2009,Huang:2010ph} and protein-ligand binding \cite{Buch06062011,Huang11,Held2011701}.

To build an MSM, one typically runs numerous molecular dynamics (MD) simulations.  A geometric criterion (e.g. RMSD) is then used to cluster the sampled conformations into a large number of microstates and the transition probabilities between pairs of states are calculated from the number of transitions observed between them \cite{Bowman_methods}.  One assumption is that conformations within each microstate are structurally similar enough to ensure fast kinetic transitions between them.  This normally requires a large number of microstates (on the order of $10,000$) even for small proteins \cite{Bowman:2009bh}.  Unfortunately, poor statistics resulting from finite sampling can result in a significant number of microstates where few transitions into/out of the state are observed, giving the false appearance that the micro state is separated from others by a large free energy barrier.   Moreover, while such microstate models are often ideal for making a direct connection to experiments because of their structural and temporal resolution, extracting human intuition from them is difficult because of the large number of states \cite{Bowman:2009bh}.

To facilitate human understanding, one typically lumps kinetically related microstates together to form a smaller number of macrostates. This coarse-graining is typically achieved using spectral methods (e.g. Perron Cluster Cluster Analysis (PCCA) algorithm \cite{Weber_PCCA,Schutte98}). 
The basic idea behind spectral methods is that metastability causes microstate transition probability matrices to have nearly block diagonal structures.  
As a result, the eigenvectors corresponding to the largest (slowest) eigenvalues are nearly piecewise constant and can be used to separate metastable states.
Poorly sampled microstates will tend to appear in the top eigenvectors as nearly constants and, as a result,  the spectral methods will separate them into their own macrostates.   
In the extreme case, poor statistics can even result in sparsely populated states that act as sources or sinks because transitions both in and out of them were not observed.  
Such sources and sinks may destroy a model's ability to predict long timescale dynamics.  In order to alleviate this problem, Noe $et$ $al.$ \cite{Noe09} removed microstates that were weakly connected to the rest, while Bowman $et$ $al.$ \cite{Bowman:2009bh} suggested subsampling the MD conformations to reduce the impact of outliers. However, both approaches require \emph{ad hoc} choices on the part of the model builder.

A number of methods have now been proposed to deal with these issues \cite{Yao_JCP09,bace,stockCG,rainsCG}.
For example, we previously presented the super-level-set hierarchical clustering (SHC) algorithm \cite{Yao_JCP09}, which was inspired by recent developments in topological model construction.
While this approach gave promising results \cite{Huang:2010ph}, this approach relied on estimates of the densities of different microstates, which are unreliable due to the high dimensionality of biomolecule's conformational spaces.

Here, we present a hierarchical formalism based on the Nystr\"{o}m extension, an algebraic principle in matrix approximation, to rigorously treat sparsely populated states with poor statistics.  
We show that the removal of sparsely populated microstates will lead to a stable approximation of the top eigenvectors of the micro state transition probability matrix that are associated with macrostates.  
Furthermore, we use the Nystr\"{o}m method to obtain an efficient lumping of microstates including those with poor statistics.  
To achieve this, we divided the microstates into multilevel subsets from high to low populations to capture the multiscale nature of biomolecule's free energy landscapes.
We then apply spectral clustering to each level set. Such a method allows us to identify the dominant metastable conformational states with significant populations.  Finally, we demonstrate the power of this new scheme on both the alanine-dipeptide and the trpzip2 peptide system.

\section{Methods}
\subsection{Kinetic Lumping with Spectral Methods}
To construct a Markov State Model (MSM) that precisely describes the conformational dynamics of a system,
one typically needs to exploit both geometric and kinetic information from MD simulations. 
Geometric distances, often measured with the RMSD distance
between configurations, can be used to identify conformations 
that can likely interconvert quickly because of their close geometric proximity. 
For example, $k$-medoids \cite{Chodera_JCP2007,Noe09} or $k$-centers \cite{Bowman_methods,Zhao_JCC2012} methods have been used to find kinetically-relevant clusterings of MD data.
This geometric clustering is often followed by kinetic lumping into macrostates, typically by spectral clustering \cite{Weber_PCCA,Schutte98}.  As discussed in the Introduction, a common issue caused by direct application of spectral method is that one captures many spurious macrostates consisting of a single, poorly sampled microstate. In order to further elaborate this issue, we first review the principles of kinetic lumping with spectral methods.

Kinetic lumping aims to group together microstates that can interconvert rapidly.  In an extreme case where metastable regions are separated by infinitely high free energy barriers, transition probabilities between these metastable macrostates will be zero.  Therefore, the transition probability matrix $P$ (a row Markov matrix satisfying $\sum_j P_{ij} = 1$ with $P_{ij}\geq 0$) contains \emph{uncoupled} Markov chains.  Let $X$ be the configuration space, $Y$ be the microstate space and $Z$ be the macrostate space. If $P$ is uncoupled with respect to the macrostate partition $\{Z_1, \ldots, Z_M\}$, $P$ can be rewritten as a block diagonal matrix after a permutation of the state indices $\{Z_1,\ldots,Z_M\}$.
\[
P=\left[
\begin{array}{cccc}
P_1 & 0 &...  & 0 \\
0 & P_2& ... & 0 \\
0 & &... &  0 \\
0 & ... & 0 & P_M
\end{array}
\right].
\]
In this case, there are $M$ (right) eigenvectors which are piecewise constants over $Z_i$ ($i=1,\ldots,M$) and associated with the maximal eigenvalue, $1$.

In biological macromolecules, metastable regions of the free energy landscape are normally separated by barriers greater than a $kT$ (i.e. thermal fluctuations).  In this case, we can view the new transition matrix $\tilde{P}$ as \emph{nearly uncoupled} Markov chains, i.e. a perturbed transition matrix $\tilde{P} = P + E$ where $P$ is uncoupled and $E$ is a matrix with small norm, denoted by $\epsilon = \|E\|$. For small enough $\epsilon$, $\tilde{P}$ has $M$ eigenvalues $\lambda_i= 1-O(\epsilon)$ ($i=1,\ldots,M$) whose eigenvectors are nearly piecewise constants, $v_i(y) = \sum_{i=1}^M c_i 1_{Z_i}(y) + O(\epsilon)$. Nearly uncoupled Markov chains are helpful to describe the conformational dynamics with a separation of timescales, where dynamics within each block $\tilde{P}_k$ are fast compared to slower dynamics between different blocks $\tilde{P}_{kl}$.

The key idea in kinetic lumping is to study the leading \emph{right} eigenvectors of the row Markov matrix $\tilde{P}=D^{-1}K$, where $K_{i,j}$ is the number of transitions from microstate $i$ to $j$ observed from simulations, and $D=diag(d_i)$ where $d_i$ is the total number of transitions starting from microstate $i$. The nearly piecewise constant structures of the right eigenvectors of $\tilde{P}$ leads to the partition of metastable macrostates. In practice, the sign structures of the right eigenvectors are often used for the lumping to construct MSMs \cite{Schutte98}.

A major issue associated with this spectral method is that it captures many spurious macrostates with low population (small $d_i$). Since microstates must have small RMSD diameters to ensure kinetic similarity within the state, a significant number of microstates will have low populations and be nearly disconnected from the rest of the state space due to insufficient sampling.  These noisy states often appear as being kinetically distinct in the top right eigenvectors of the transition matrix. To remove such noise from the signal, regularization must be applied. Some heuristic approaches have been adopted in the literature, like leaving-out certain rarely visited states \cite{Noe09} or subsampling the MD data \cite{Bowman:2009bh}. However, how much data to discard remains an open question in such \emph{ad hoc} methods.

In the next section, we will introduce the Nystr\"{o}m Method, an algebraic principle for matrix approximation by sub-rows (columns) that provides a mathematical foundation for the heuristic methods described above and, moreover, points to a new technique for coarse-graining MSMs.

\subsection{Nystr\"{o}m Method}
Let us define a transition count matrix $K$ whose elements ($K(i,j)$) denote the number of transitions observed from state $i$ to state $j$ for all $i,j\in Y$. 
We will expect that the transition probability matrix $\tilde{P}=D^{-1}K$ with $D=diag(d_{i})$ where $d_i=\sum_{j} K(i,j)$, is a nearly uncoupled Markov matrix. 
Unfortunately when there are not sufficient samples, there are often spurious macrostates with low population which contribute to the nearly uncoupled dynamics in $\tilde{P}$.  In the following we will suggest a method based on the Nystr\"{o}m method to eliminate these noisy states.

\subsubsection{Application to symmetric matrix}
Originally, the Nystr\"{o}m approximation was developed to find an approximate eigen-decomposition of a symmetric matrix based on its submatrices.  
We first illustrate this idea by applying it to the transition count matrix, which is symmetric due to the reversibility of molecular dynamics.  
Assume that a symmetric and non-negative transition count matrix $K$ is partitioned by

\begin{equation}
K = \left[
\begin{array}{cc}
A & B \\
B^T & C
\end{array}
\right].
\end{equation}

Let $ A = U \Lambda U^T$ be the eigen-decomposition of $A$, then we can define the Nystr\"{o}m approximation of $K$ to be
\begin{equation}
\hat{K} = \left[
\begin{array}{c}
U  \\
B^T U\Lambda^{-1}
\end{array}
\right]
\Lambda
\left[
\begin{array}{cc}
U^T  & \Lambda^{-1} U^T B
\end{array}
\right] = \left[
\begin{array}{cc}
A & B \\
B^T & B^T A^{-1} B
\end{array}
\right]
\end{equation}
Therefore $\| K - \hat{K} \| = \| C - B^T A^{-1} B\|\approx 0 $ if the entries of $C$ and $B$ are far smaller than those of $A$. 
In this case, $\hat{U}^T=[U^T, \Lambda^{-1} U^T B]$ provides a good approximation of the leading eigenvectors of $\hat{K}$ and the original $K$. The sign structure of $U$ decided by $A$ will be stable compared to $K$.

\subsubsection{Application to transition probability matrix.}
To construct MSMs, we need to deal with the non-symmetric transition probability matrix $\tilde{P}$. Our purpose is to find the right eigenvectors $u$ of $\tilde{P}$, s.t. $\tilde{P} u =\lambda u$, which
will be used to find metastable states. 
Recall that $\tilde{P}=D^{-1} K$.
Therefore, $\tilde{P} u = D^{-1} K u = \lambda u \Leftrightarrow D^{-1/2} K D^{-1/2} v = \lambda v$ where $ v =D^{1/2} u$. 
So, to find the eigenvectors of $\tilde{P}$, it suffices to find the eigenvectors $v$ of the symmetric matrix $K_n = D^{-1/2} K D^{-1/2}$, through which the right eigenvectors of $\tilde{P}$ can be obtained by $u=D^{-1/2} v$. To find a robust coarse-graining into metastable states, the Nystr\"{o}m method can be extended to treat $K_n$.

Consider $K_n$ partitioned in the following way,
\begin{equation}
K_n= D^{-1/2} K D^{-1/2} = \left[
\begin{array}{cc}
D_A & 0 \\
0 & D_B
\end{array}
\right]^{-1/2}
\cdot
\left[
\begin{array}{cc}
A & B \\
B^T & C
\end{array}
\right]
\cdot
\left[
\begin{array}{cc}
D_A & 0 \\
0 & D_B
\end{array}
\right]^{-1/2}
\end{equation}
where $D = diag(d_{i})=diag(D_A,D_B)$, and $d_{i} = \sum_{j} K_{ij}$.

Define $\hat{D}_A = diag(\sum_j A_{ij}) $ and eigen-decomposition $\hat{D}_A^{-1/2} A \hat{D}_A^{-1/2} = V \Lambda V^T $.  Then $\hat{D}_A^{-1} A U = U \Lambda $ where $U=\hat{D}_A^{-1/2} V$ and $U^T \hat{D}_A U = I$.
Define $\hat{D}_B = diag(\sum_j B_{ij})$. Nystr\"{o}m extension for $K_n$ is
\begin{eqnarray*}
\hat{K}_n & = & \left[
\begin{array}{c}
V  \\
\hat{D}_B^{-1/2} B^T \hat{D}_A^{-1/2} V\Lambda^{-1}
\end{array}
\right]
\Lambda
\left[
\begin{array}{cc}
V^T  & \Lambda^{-1} V^T \hat{D}_A^{-1/2} B \hat{D}_B^{-1/2}
\end{array}
\right] \\
& = & \left[
\begin{array}{cc}
A_n & B_n \\
B_n^T & B_n^T  A_n^{-1} B_n
\end{array}
\right]
\end{eqnarray*}
where $A_n = \hat{D}_A^{-1/2} A \hat{D}_A^{-1/2}$, $B_n = \hat{D}_A^{-1/2} B\hat{D}_B^{-1/2}$.  If we concentrate on the major submatrix $A$,  $A_n$ has eigen decomposition $A_n= V \Lambda V^T $, and thus $U=\hat{D}_A^{-1/2} V$ is the right eigenvector of $P_A = \hat{D}_A^{-1}A$. 
The Nystr\"{o}m method therefore approximates the right eigenvectors of $\tilde{P}=D^{-1}K$ as $[U^T, (\hat{D}^{-1}_B B^T U\Lambda^{-1})^T]^T$.

It is clear that in the Nystr\"{o}m method, $\hat{K}_n$ approximates $K_n$ well if $C,B<<A$, whence $D_A\approx \hat{D}_A$, $D_B\approx \hat{D}_B$, and $D_B^{-1/2} C D_B^{-1/2} - B_n^T A_n^{-1} B_n \approx 0$. To ensure that the entries of $C$ and $B$ are as small as possible compared to those of $A$, one can either assign all the sparse (or rare) states to the block $C$ or randomly down-sample the conformations such that rare states are not included in the block $A$. 
These two schemes correspond to the heuristic treatments by Noe $et$ $al.$ \cite{Noe09} and Bowman $et$ $al. $\cite{{Bowman:2009bh}}, respectively. 

In this paper we use the following greedy method to construct the submatrix $A$ deterministically. 
We first sort the rows (columns) of $K$ in decreasing order according to the row (column) sum.
We then choose $A$ to be the $m$-by-$m$ matrix consisting of the $m$ largest rows and columns. 
In practice, one often observes that the distribution of microstate populations is roughly a power law distribution, which implies $B$ collects exponentially many states/rows with exponentially small populations.

\subsubsection{Recovery of original Markov Chains.}

In the Nystr\"{o}m  approximation discussed above, the eigenvectors of the submatrix $A$ have the same sign structures as the right eigenvectors of the transition probability matrix if the entries of $A$ are sufficiently large compared to those of $B$ and $C$. However, in order to recover the original Markov chains using Nystr\"{o}m method, a few other conditions need to be satisfied.  Let's consider that the original transition probability matrix  $\tilde{P}=D^{-1} K$ can be partitioned into $M$ macrostates: $\mathcal{M}=\{Z_1,\ldots,Z_M\}$.  If $\tilde{P}$ is an uncoupled Markov chain with $M$ piecewise constant eigenvectors associated with eigenvalue $1$, the partition $\mathcal{M}$ can be exactly recovered under the following three conditions. First, at least one microstate in any macrostate $Z_i$ has to appear in the submatrix $A$, otherwise there is no possibility to reconstruct it.  
Second, any macrostate $Z_i$ must not break into disconnected parts in $A$.  This condition ensures that downsampling in the Nystr\"{o}m  approximation does not generate incorrectly break one macrostate into multiple pieces. Finally, every rare microstate with indices in $B$ should be directly connected to one or more large microstates with indices in $A$. With these conditions, one can recover such macrostates by the Nystr\"{o}m method, merely from submatrices $A$ and $B$, via the following steps.

\begin{enumerate}
\item Find eigen-decomposition of $A_n = \hat{D}_A^{-1/2} A \hat{D}_A^{-1/2} =  V \Lambda V^T $;
\item Construct $U=\hat{D}_A^{-1/2}V$, find piecewise constant vectors in column vectors of $U$ and the partition $\{\hat{Z}_1, \ldots, \hat{Z}_M\}$ associated with them;
\item Assign each rare state $k$ to the partition it has the maximal transition probability to, $\max_i \tilde{P}_{k \hat{Z}_i}=\max_i \sum_{j\in Z_i} \tilde{P}_{kj}$.
\end{enumerate}

These conditions can be summarized precisely in Theorem A (see Appendix).


However, in molecular dynamics simulations of biological macromolecules, we normally deal with nearly uncoupled Markov Chains determined by the underlying metastable regions of the free energy landscape.  In this case, we can  introduce a small perturbation on the uncoupled Markov Chain discussed above: $A=A_0 + E$ with the block-diagonal $A_0$ for the uncoupled Markov Chain.  The partition $\tilde{P}=\{Z_1,\ldots,Z_M\}$ can be recovered using Nystr\"{o}m  method if the following three conditions can be satisfied.  The first condition is identical with the uncoupled Markov Chain case and requires that some microstates of $Z_i$ are selected from $A$.  In the second condition, any two microstates in $Z_i$ can always be connected directly or through other microstates in $A$ regardless of the magnitude of the perturbation $\epsilon=\|E\|$.  In other words,  $Z_i$ does not break into disconnected parts in $A$ with small enough perturbation $\epsilon$.  The last condition requires that every microstate in $B$ should be always directly connected to one or more microstates in  $A$ regardless of the size of $\epsilon$.  In practice, a sufficiently long lag time may be used to obtain the transition probability matrix in order to satisfy this condition, which can be implemented by applying step 3 above to transition probabilities with long enough lag time.  These conditions can be summarized in Theorem B (see Appendix).


The challenge for applying the Nystr\"{o}m method is the lack of prior knowledge of how best to split the transition matrix into submatrices  $A$ and  $B$.  
Moreover, due to the multiscale nature of free energy landscapes \cite{Huang:2010ph}, this split will largely depend on the scale of the model, i.e. how coarsely we decompose the conformation space.  
Therefore, we pursue a hierarchical Nystr\"{o}m approximation by ranking all the rows (columns) by their sums (or populations) and dividing the matrix $K$ into multilevel blocks.  


\subsection{Multilevel analysis on free energy landscapes}

\begin{figure}[!h]
\begin{center}
\includegraphics[width=4in]{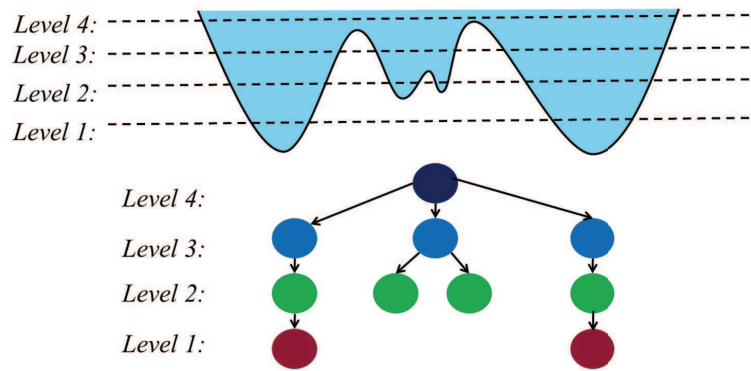}
\end{center}
\caption{
Schematic figure for the multilevel analysis of a free energy landscape. (a) a 1-D free energy landscape divided into four levels; (b) a cluster tree representing this free energy landscape. At level one, two nodes are formed that correspond to the two deepest free energy minima. At level two, four nodes are identified for four free energy minima. At level three and four, the number of nodes are reduced since some free energy minima are connected.}
\label{fig:densitytree}
\end{figure}

The free energy landscapes of biomolecules are rugged, having a large number of local minima separated by barriers with various height. Therefore the metastable macrostates have a
multiscale nature depending on the energy barriers and equivalently the intrinsic timescales. Such a multiscale landscape can be
accurately represented by a cluster tree (Figure ~\ref{fig:densitytree}) \cite{Huang:2010ph}.

Given a free energy landscape, construction of such cluster trees can be performed as follows. (a) Divide the conformation space into overlapping level sets according to free energy, so that each level contains all previous levels with lower free energy. (b) For each level, apply spectral clustering to the largest connected component. (c) A cluster tree is generated by connecting nodes lying in adjacent levels that have nonempty intersection.

An example is given in Figure~\ref{fig:densitytree}.
In the coarsest level, two metastable states (corresponding to the two nodes belonging to level 1 in Figure~\ref{fig:densitytree}) are desired with a large implied time scale for transitioning between them. In a less coarse model, the system may exhibit four metastable states (corresponding to the four nodes belonging to level 2 in Figure~\ref{fig:densitytree}), and two of them are crossed with relatively faster speed. To capture metastable states, one has to keep in mind such a multiscale nature.

The challenge in conformation dynamics lies in that we do not know the underlying free-energy landscape. Even though the free energy will decide the sample distribution in conformation space, it is impossible to reach an accurate estimation of densities in such a high dimensional space. Therefore it is impossible to pursue an accurate multilevel analysis of such free energy landscapes. 
Fortunately the Nystr\"{o}m method leads us to a reasonable multilevel approximation scheme.

Microstates with high populations are visited frequently and, therefore, tend to be highly correlated with metastable states or free energy basins. On the other hand, microstates with low population are only visited rarely, suggesting they lie on energy barrier. 
Therefore we may apply the Nystr\"{o}m method progressively from highly populated states to rarely visited states, which leads to a natural way to explore free energy landscapes according to their multiscale nature. The following algorithm works efficiently in the application examples shown in the Results section.

\begin{algorithm}[H]
\caption{Hierarchical Nystr\"{o}m Extension Graph (HNEG) \label{HNEG}}
{\small \begin{algorithmic}
\STATE (1) Sort microstates by their population in decreasing order, and divide all microstates into $m$ level sets: top $p_1, p_2,\ldots, p_m$ populated states, where $p_1\leq p_2\leq \ldots \leq p_m$;
\STATE (2) On each level, perform spectral clustering only with top $p_j$ ($1\leq j \leq m$) microstates,
\STATE (3) Draw a Hierarchical Nystr\"{o}m Extension Graph: \\
\quad  A. Node set: for each level, each cluster/group is represented by a node; \\
\quad  B. Edge set: an edge is added if the associated two nodes are in adjacent levels with nonempty intersections; \\
\quad C. Gradient flow arrows: for each edge, an arrow is added pointing from high to low level; \\
\STATE (4) Lumping: \\
\quad A. Set each attraction node (with zero out-degree) as metastable states; \\
\quad B. Expand those metastable states to its associated attraction basins (the intermediate nodes which flow down to single attraction nodes along gradient flows); \\
\quad C. Assign microstates on barrier nodes (those branching nodes which flow down to more than one attraction nodes) to the attraction basin with maximal transition probability.
\end{algorithmic}}
\end{algorithm}

Unlike the case of a cluster tree built from a known free-energy landscape, the hierarchical Nystr\"{o}m method for nearly uncoupled Markov chains generally constructs a graph instead of a tree. Note that in this algorithm, the most important free parameter is how to divide the microstates into level sets, \emph{i.e.} the choice of the $p_j$. In this study, we have applied the Bayes factor approach introduced by Bacallado \emph{et. al.} \cite{Sergio09-JCP} to select the optimal level set.  A Bayes factor ($B(L_1,L_2)=P(L_1\mid D)/ P(L_2 \mid D)$) is the ratio of posterior probabilities for two lumpings ($L_1$ and $L_2$) given finite sampling and a common set of microstates (D).  We first generate a large number of models by systematically varying the level sets, and then select the optimal level set as that can produce the model with the highest posterior probability.  The second important free parameter is how to choose the number of clusters during spectral clustering within each level set. In this work, the number of clusters will be chosen based on the maximal spectral gap in top eigenvalues as in PCCA \cite{Weber_PCCA,Schutte98}.

\subsection{Simulation Details}

\textbf{Alanine Dipeptide} The alanine dipeptide dataset is taken from Chodera $et$ $al.$ \cite{Chodera_JCP2007}.  It consists of  975 20-ps NVE simulations with conformations stored every 0.1 ps, for a total of 195k conformations.  Please refer to \cite{Chodera_JCP2007} for additional details on the simulations.  We split the conformations into 5000 microstates using a K-centers clustering algorithm \cite{Bowman_methods}. The distance between a pair of conformations is determined by their backbone root mean square deviation (RMSD).

\textbf{Trpzip2} The trpzip2 simulation dataset is taken from Zhuang $et$ $al.$ \cite{Zhuang_JPC2011}.  It contains 830 50-ns MD simulations with conformations saved every 100ps.  This results in a total of 415k conformations.  Please refer to \cite{Zhuang_JPC2011} for additional details on the simulations.  We have performed K-centers clustering \cite{Bowman_methods} to divide the conformations into 2000 clusters based on the heavy atom RMSD.

\section{Results}

\subsection{Alanine-dipeptide}

\begin{figure}[!ht]
\begin{center}
\includegraphics[width=4in]{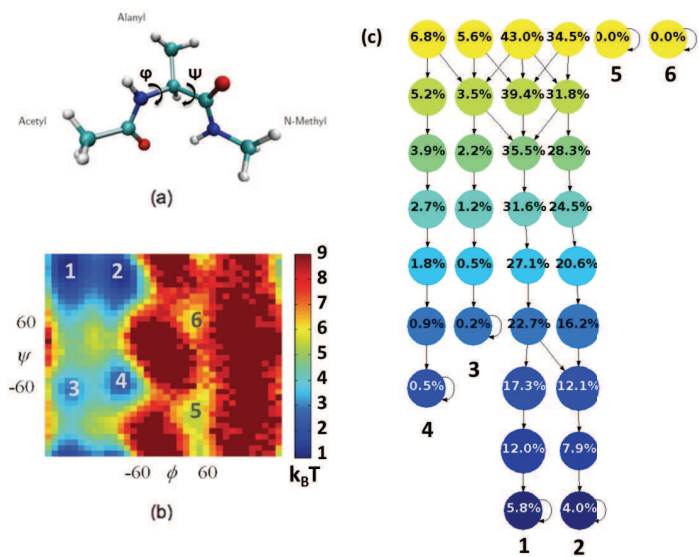}
\end{center}
\caption{
An illustration of Hierarchical Nystr\"{o}m Extension Graph (HNEG) applied to the alanine dipeptide system. (a) A conformation of the alanine dipeptide with two torsion angels ($\phi$ and $\psi$) labeled. (b) The free energy landscape projected onto the $\phi-\psi$ plane, where the red color indicates regions of high density or low free energy. (c) A Hierarchical Nystr\"{o}m Extension Graph containing 9 levels constructed for this system.
}
\label{fig:ala}
\end{figure}

To validate the algorithm developed here, we first apply it to a simple alanine dipeptide system.  
For alanine dipeptide, it is easy to obtain equilibrium sampling and an accurate representation of the free energy landscape by projecting it onto a pair of torsion angles: $\phi$ and $\psi$  (see Fig. ~\ref{fig:ala}a). Therefore we can visualize the resulting states and check their various properties on this projection of the free energy landscape.  
Such projection is displayed in Fig ~\ref{fig:ala}b, and the free energy is calculated by $W=-kT\ln (P_i/P_0)$, where $P_i$ is the number of conformations that lie in bin $i$ divided by the total of number conformations, and $P_0$ is a constant. 
We discretize the torsion plane by dividing it into square bins of 10$^\circ$ by 10$^\circ$.  
As shown in Fig ~\ref{fig:ala}b, there are six minima or metastable states centered at around: (-140$^\circ$, 160$^\circ$), (-60$^\circ$,140$^\circ$), (-140$^\circ$, 160$^\circ$), (-60$^\circ$, -50$^\circ$), (50$^\circ$, -100$^\circ$), and (40$^\circ$, 60$^\circ$).

In order to generate the Hierarchical Nystr\"{o}m Extension Graph (HNEG) to describe the conformational dynamics of the alanine dipeptide, we first applied the k-centers clustering algorithm to divide the conformations into 5000 microstates.  
Clusters generated from the k-centers clustering algorithm have an approximately uniform distance to their cluster centers, and thus result in a strong correlation between the cluster population and its density \cite{Zhao_JCC2012,Bowman_methods,Huang:2010ph}. 
We then followed the procedure introduced in Section IIC to construct the hierarchical directed graph using the Nystr\"{o}m Expansion method.  
One example of a HNEG is shown in Fig. ~\ref{fig:ala}c. This graph is generated using the level set $[0.1, 0.2, 0.3,\ldots, 0.8, 0.9]$, and arrows are used to connect intersecting nodes in the adjacent levels as gradient flows.  There are six attraction nodes (labeled from 1-6) each lying in to a different free energy basin as shown in Fig ~\ref{fig:ala}c.  Attraction nodes 1 and 2 appear in the first level (top 10\% population), indicating the deepest free energy basins. While attraction nodes 5 and 6 belong to the last level and correspond to the shallowest free energy basins. These results are consistent with projection of the free energy landscape shown in Fig. ~\ref{fig:ala}b.  Therefore, the HNEG is able to provide a hierarchical representation of the free energy landscape.  Moreover, starting from the 6 attraction nodes, we can further lump all the microstates into six metastable states following {\bf{Algorithm 1}} in Sec IIC.

To compare the performance of our lumping algorithm with the popular PCCA method, here we apply the Bayes factor approach \cite{Sergio09-JCP}.  
As discussed in Sec IIC, the Bayes factor compares the posterior probabilities of two lumpings constructed from a common microstate set, and the larger the Bayes factor is, the more comparative advantage the first lumping has. For example $\ln B = 50$ implies that $P(L_1\mid D)/P(L_2 \mid D)= e^{50} \approx 5.18\times 10^{21}$, whence lumping $L_1$ is much better than $L_2$.  Since MSMs generated by our algorithm is not unique and depend on the construction of level sets, we have generated a number of different MSMs using a systematic variation of the level sets to compare with the PCCA method.

The level sets are mainly determined by two factors: the population cut-off (i.e. microstates to be included in the submatrix $A$ in the Nystr\"{o}m method) and the number of levels.  We also noticed that the largest microstate has a population of ~6\%, so we select the 1st level to be at 10\% to ensure a sufficient number of states in each level.  Specifically, we have selected 9 different population cut-offs ($p_m$): 20\%, 30\%, \ldots, 90\% and 99\% and 10 different numbers of levels (from 6 to 15). For example, the combination of 90\% cut-off and 9 levels will generate a level set of $[0.1, 0.2, 0.3, \ldots, 0.8, 0.9]$.  Altogether, we have obtained 90 different level sets for MSM construction.

\begin{figure}[!h]
\begin{center}
\includegraphics[width=0.8\textwidth]{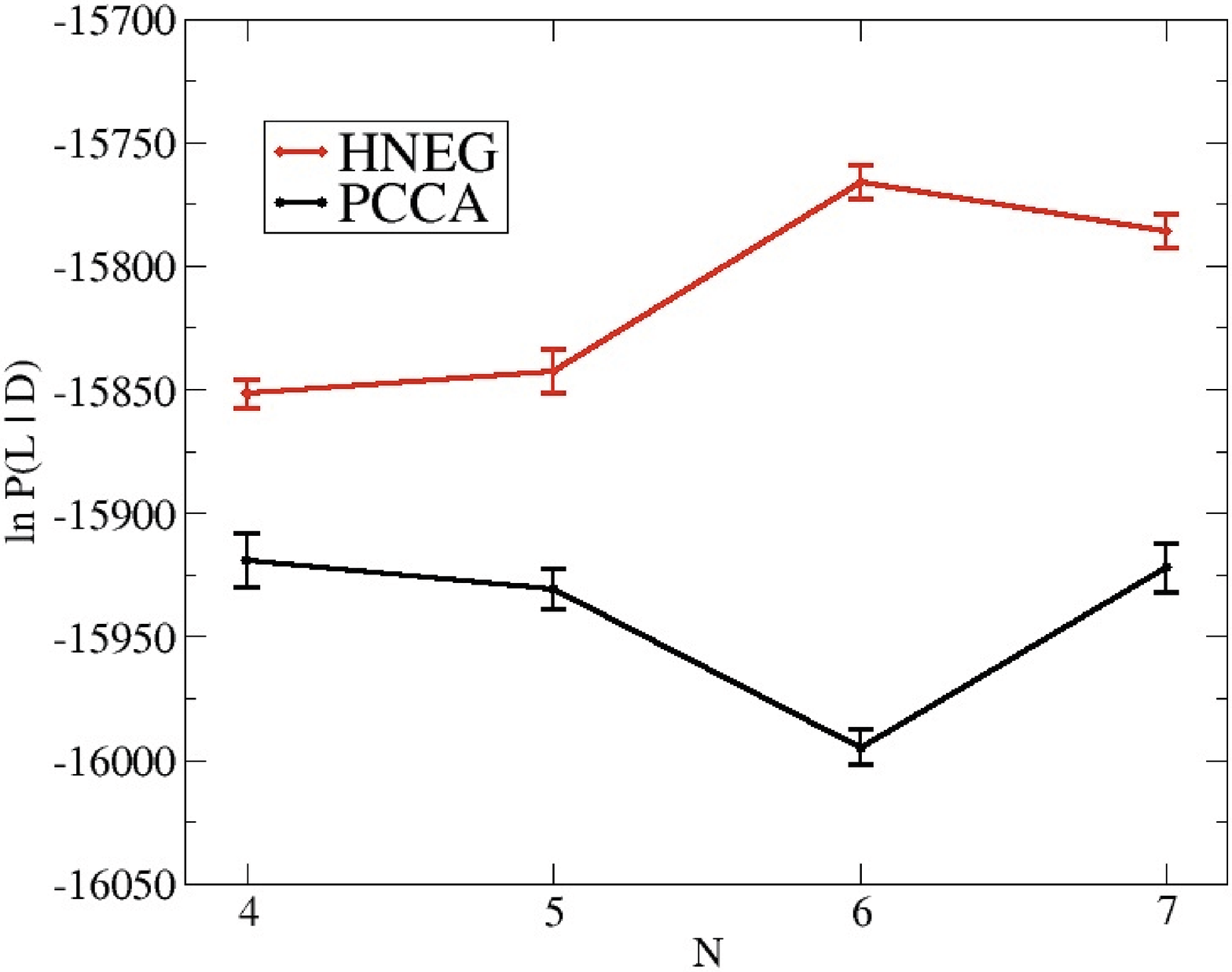}
\end{center}
\caption{
Bayesian comparison of MSMs constructed by the Hierarchical Nystr\"{o}m Extension Graph (HNEG) and the PCCA method for the alanine dipeptide.  The y-axis displays the logarithm of the posterior probability ($\ln (P(L_1\mid D))$) for models generated by HNEG (red) and PCCA(black). The logarithmic Bayes factor $\ln B = \ln (P(L_1\mid D))_{HNEG}-\ln (P(L_2\mid D)_{PCCA})\gtrsim 100$, indicating that HNEG consistently provides a better lumping than PCCA.
}
\label{fig:alab}
\end{figure}

The Bayes factor test shows that the models generated by our algorithm have a consistently higher posterior probability ($\ln B > 100$) than those obtained from the PCCA method (see Fig.~\ref{fig:alab}).  Moreover, the models with six metastable states have the highest Bayes factors, indicating that a 6-state model is an optimal choice, which is consistent with the earlier study \cite{Chodera_JCP2007}.

\subsection{Trpzip2}

In this section, we have further applied our algorithm on a larger system: a 12-residue $\beta$-hairpin trpzip2 peptide.  Its folding has been extensively studies by both experimental and computational techniques \cite{Cochran08052001, Zhuang_JPC2011,Chodera_JCP2007,Cui_CurrPChem2012,doi:10.1021/jp104017h,doi:10.1021/ja0493751,BIP:BIP20101}. In particular, a few recent studies have constructed MSMs from MD simulations to identify the metastable states during its folding process \cite{Zhuang_JPC2011,Chodera_JCP2007}.

\begin{figure}[!h]
\begin{center}
\includegraphics[width=0.9\textwidth]{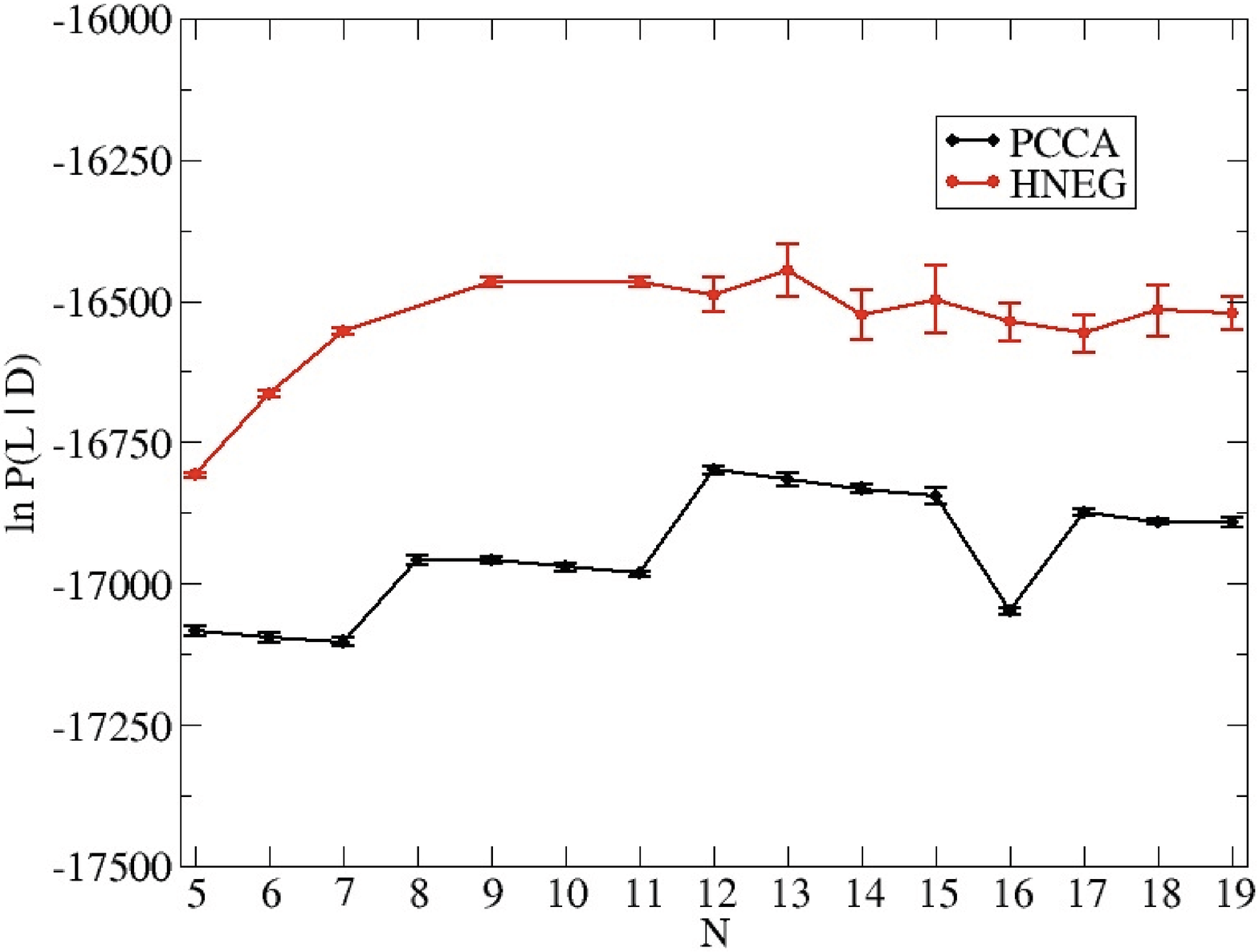}
\end{center}
\caption{
Bayesian comparison of MSMs constructed by the Hierarchical Nystr\"{o}m Extension Graph (HNEG) and the PCCA method for the trpzip2 peptide. The y-axis displays the logarithm of the posterior probability ($\ln (P(L_1\mid D))$) for models generated by HNEG (red) and PCCA(black). The logarithmic Bayes factor $\ln B = \ln (P(L_1\mid D))_{HNEG}-\ln (P(L_2\mid D)_{PCCA})\in [250,500]$, indicating that HNEG consistently provides a better lumping than PCCA for the trpzip2 system.}
\label{fig:trpzipa}
\end{figure}

The Bayes comparison results clearly demonstrate the advantage of our algorithm over the PCCA method.  As with the alanine dipeptide system, we have generated a number of different lumpings by varying the level sets of the 2000 microstates generated by K-centers clustering.  Specifically, we select 13 different population cut-offs ($p_m$ =41\%, 45\%, 50\%, 55\%,\ldots, 90\%, 95\%, 99\%) and 20 different numbers of levels (m=11,...,30), resulting in a total of 260 different models.  As shown in Fig.~\ref{fig:trpzipa}, all 260 models have substantially higher posterior probabilities than those obtained from the PCCA method.  Furthermore, models with 13 metastable macrostates have the highest posterior probabilities, which are consistent with our previous study \cite{Zhuang_JPC2011}.  Therefore, we select the model with the highest posterior probability as the optimal lumping.
In Fig. 5, we have displayed the representative structures from each of the macrostates of the optimal lumping together with its equilibrium population obtained from the MSM.  The macrostate 3 has the highest population (around 38\%) and corresponds to a folded hairpin structure, indicating that the trpzip2 peptide still contains a large fraction of the hairpin component at 350K.

Our results also show that the new algorithm efficiently captures highly populated metastable regions of the transition probability matrix, while the PCCA method  first separates rarely visited sparse states.  To visualize this, we have compared the nearly diagonal block structures of the transition probability matrices of a representative lumping obtained by our algorithm and  PCCA (see Fig.~\ref{fig:trpzipc}a).   These results show that PCCA has the tendency to first separate sparsely populated states as metastable states (6 out of 11 states have populations $<2\%$), while our algorithm efficiently identifies the highly populated metastable regions.  Among the 260 models generated by our algorithm, the number of large states ($>2\%$) converges to ~9 with a total of less than 20 metastable states (see Fig.~\ref{fig:trpzipc}b).  However, PCCA needs models with nearly 80 states in order to identify the same number of large metastable states.

Our results also demonstrate that the identification of the major metastable states using the Nystr\"{o}m method is robust to leaving out different fraction of rarely visited microstates.  
After including $>50\%$ of the most populated microstates, our algorithm consistently identifies the same number of large macrostates (with population $>1\%$, 2\% or 3\%) (see Fig.~\ref{fig:trpzipd}).  
In order to further check whether our algorithm can consistently identify large macrostates containing similar conformations in addition to the same populations, we have performed pairwise mutual information calculations between the optimal lumping and all other lumpings. For a pair of two lumpings ($f$ and $g$), the mutual information is defined as
\begin{equation}
\ I(f, g) = \sum_{i,j} P(f(x) = i, g(x)=j) \log \frac{P(f(x) = i, g(x)=j)}{P(f(x) = i) P_X(g(x) = j)},
\label{equ:mutual}
\end{equation}
where $P(f(x) = i)$ is the probability for microstate $x$ belonging to macrostate $i$ in the lumping $f$, and $P(f(x) = i, g(x)=j)$ is the joint probability for microstate $x$ belonging to both macrostate $i$ in the lumping $f$ and macrostate $j$ in the lumping $g$. 

In Fig.~\ref{fig:trpzipf}, we have plotted the histogram of pairwise mutual information between the optimal lumping (with thirteen macrostates) and all other lumpings generated by our algorithm (259 of them).  All these lumpings from our method have an average mutual information of $2.104$.  
This value is close to its theoretical upper limit (the entropy of the optimal lumping, $3.313$), and well above that of the model produced by the PCCA method (with thirteen macrostates and the mutual information of $0.785$).  
For better comparison, we have also shown that the mutual information between random lumpings and the optimal lumping is only around $0.0513$.  
These results clearly demonstrate that models generated by our Nystr\"{o}m method with different level sets consistently have a larger similarity with the optimal lumping compared to the PCCA method.

In order to further illustrate that different lumpings generated from our method contain a common set of large macrostates that share similar protein conformations or microstates,  we have plotted the the joint probability $P(f(x) = i, g(x)=j)$ between the optimal lumping and a representative lumping (see Equ.~\ref{equ:mutual} for the detailed definition).  We chose the representative lumping to have a mutual information close to the mean value of all the lumpings generated by our algorithm.  Both the optimal lumping and the representative lumping contain 9 large macrostates (with population $>2\%$).  The joint probability matrix is thus a 9 by 9 matrix (see Fig.~\ref{fig:trpzipe} ), and each matrix element represents the overlapping of the microstates in a pair of macrostates each belonging to a different lumping.   
After a permutation, it is clear that large macrostates in two compared lumpings contain a large fraction of identical microstates (with large diagonal matrix elements but small off-diagonal elements).

\section{Conclusion and Discussion}

We have developed an algorithm based on the Nystr\"{o}m method and its multilevel extensions to construct MSMs from Molecular Dynamics simulation trajectories. 
Our algorithm is shown to be efficient in eliminating noise due to insufficient sampling by focusing on the dominant submatrix of the transition probability matrix.  
Therefore, the Hierarchical Nystr\"{o}m Extension Graph (HNEG) method can greatly improve over other popular kinetic clustering algorithms such as PCCA, which tends to identify sparse states that are very weakly coupled to the rest of phase space before identifying real metastable states in denser regions of phase space. 
We have demonstrated that HNEG can generate a number of MSMs at different resolutions.  

In practice, HNEG has a similar protocol to Super-level-set Hierarchical Clustering (SHC) which was developed by the authors as a prototype before \cite{Huang:2010ph}.  SHC relies on the rough estimation of the conformation density using the populations of microstates from K-centers clustering.  However, estimating densities in high dimensional spaces is challenging, and the K-centers algorithm only generates clusters with approximately equal radii, so small variances in the cluster radius may induce large volume differences. On the other hand, the algebraic framework based on the Nystr\"{o}m method does not rely on density estimation and leads to efficient algorithms to pursue metastable macrostates with a multiscale analysis.


\begin{acknowledgments}
We acknowledge the support from the  National Basic Research Program of China (973 Program 2011CB809105, 2012CB825501 to Y.Y. and 2013CB834700 to X.H.), NSFC (61071157 to Y.Y. and 21273188 to X.H.), Microsoft Research Asia, a professorship in the Hundred Talents Program at Peking University to Y.Y., Hong Kong Research Grants Council GRF 661011, F-HK29/11T, HKUST2/CRF/10 to X.H., and University Grants Council SBI12SC01 to X.H.
\end{acknowledgments}

\newpage
\section*{References}

\newpage
\section*{Appendix}
{\textbf{Theorem A  (Exact Recovery).}} {\textit{Let $d_i=\sum_j K_{ij}$ be the degree of microstate $i$, and $P=D^{-1} K$ be an uncoupled Markov matrix with respect to partition $P=\{Z_1,\ldots,Z_M\}$. Given $\theta>0$, denote by $I_\theta=\{i: d_i \geq \theta\}$ the microstates selected for submatrix $A$ in Nystr\"{o}m method. Then Nystr\"{o}m method will exactly recover those $Z_k$, if and only if the following holds:
\begin{enumerate}
\item  $Z_k\cap I_\theta \neq \emptyset$;
\item Any two microstates in $Z_k$ will be connected by a path in $I_\theta$;
\item For any microstate of $Z_k$ outside $I_\theta$, it is directly connected to a microstate of $Z_k$ in $I_\theta$.
\end{enumerate}}}

{\textbf{Proof of Theorem A}} Since $P=D^{-1}K$ is an uncoupled Markov matrix, the submatrix $A$ is block-diagonal. The first condition and the second condition ensures that the microstates of $Z_k$ which are selected in $A$ are within the same block, whence there is a right eigenvector of $\hat{D}^{-1}_A A$, $U_k = \I_{Z_k\cap I_\theta}$ as the indicator function of $Z_k\cap I_\theta$. Now consider the Nystr\"{o}m approximation
\[ \tilde{u}_k=
\left(
\begin{array}{c}
U_k \\
D^{-1}_B B^T U_k
\end{array}
\right) =
\left(
\begin{array}{c}
\I_{Z_k\cap I_\theta} \\
D^{-1}_B B^T \I_{Z_k\cap I_\theta}
\end{array}
\right)
\]
The $j$-th row in $D^{-1}_B B^T \I_{Z_k\cap I_\theta} $ summarizes the probability that microstate $j$ which is outside $I_\theta$, to be connected in $Z_k\cap I_\theta$. If $j\notin Z_k$, by uncoupled Markov matrix assumption, such a probability is $0$; If $j\in Z_k$, such a probability could be $1$ or $0$ depending on whether $j$ is directly connected to $Z_k\cap I_\theta$, whence by the third condition, the probability must be $1$ in this case. In summary, $\tilde{u}_k$ is an indicator function on $\I_{Z_k}$ on all microstates, whence $Z_k$ is exactly recovered.

{\textbf{Theorem B (Asymptotic Recovery under Noise).}} {\textit{Let $\tilde{P}=D^{-1} K=P_0 + E$ ($\epsilon=\|E\|$) be a nearly uncoupled Markov matrix with respect to partition $\M=\{Z_1,\ldots,Z_M\}$. Given $\theta>0$, let $I_\theta=\{i: d_i \geq \theta\}$. Then Nystr\"{o}m method with submatrix selected from $I_\theta$ rows and columns will recover $Z_k$ as $\epsilon\to 0$, if and only if the following holds:
\begin{enumerate}
\item $Z_k\cap I_\theta \neq \emptyset$.
\item Any two microstates in $Z_k$ will be connected by a `highway' in $I_\theta$, in the sense that for any $i,j\in Z_k \cap I_\theta$, there is a path $(i=k_0,k_1,\ldots,k_t=j)$ in $I_\theta$ such that $\lim_{\epsilon\to 0}K_{k_{t-1},k_{t}}\geq \gamma >0$ for some constant $\gamma$;
\item Any microstate of $Z_k$ outside $I_\theta$ is directly connected to some microstates of $Z_k$ `strongly' in $I_\theta$, in the sense that for any $i\in Z_k$ ($k=1,\ldots,M$), there is $j\in Z_k\cap I_\theta$ such that $\lim_{\epsilon\to 0} K_{i,j} \geq \gamma >0$ for some constant $\gamma$.
\end{enumerate}
}}

{\textbf{Proof of Theorem B}} The proof of this theorem follows the same reasoning above. Notice that when $\tilde{P}=D^{-1} K=P_0 + E$ ($\epsilon=\|E\|$) be a nearly uncoupled Markov matrix with respect to partition $\M=\{Z_1,\ldots,Z_M\}$, one has $U_k=\I_{Z_k\cap I_\theta} + O(\epsilon)$ and $D^{-1}_B B^T U_k\lambda^{-1}_k = D^{-1}_B B^T\I_{Z_k} + O(\epsilon)$. So in the limit $\epsilon\to 0$, the three conditions recover the three conditions in noise-free case.

\newpage
\section*{Figures}
\newpage

\newpage

\newpage

\newpage

\begin{figure}[!h]
\begin{center}
\includegraphics[width=0.8\textwidth]{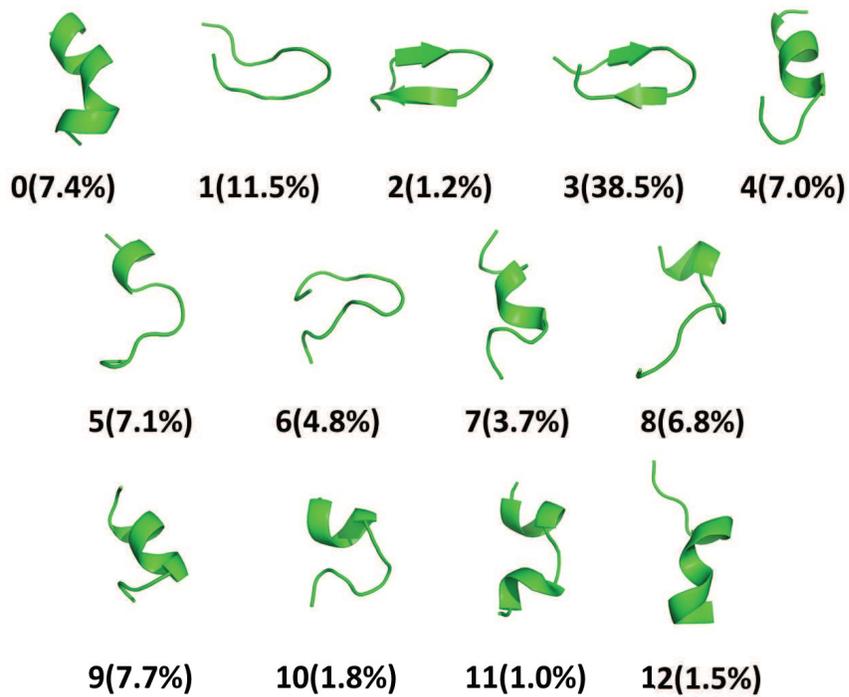}
\end{center}
\caption{
Representative structures of the 13 macrostates from the optimal lumping (with the highest posterior probability) for the trpzip2 system. Their equilibrium populations are also displayed. Macrostate 3 corresponds to a folded hairpin structure and has the largest population (38.5\%), indicating that the trpzip2 peptide still has a significant fraction of the folded structure at 350K.
}
\label{fig:trpzipb}
\end{figure}
\newpage

\begin{figure}[!h]
\begin{center}
\includegraphics[width=0.9\textwidth]{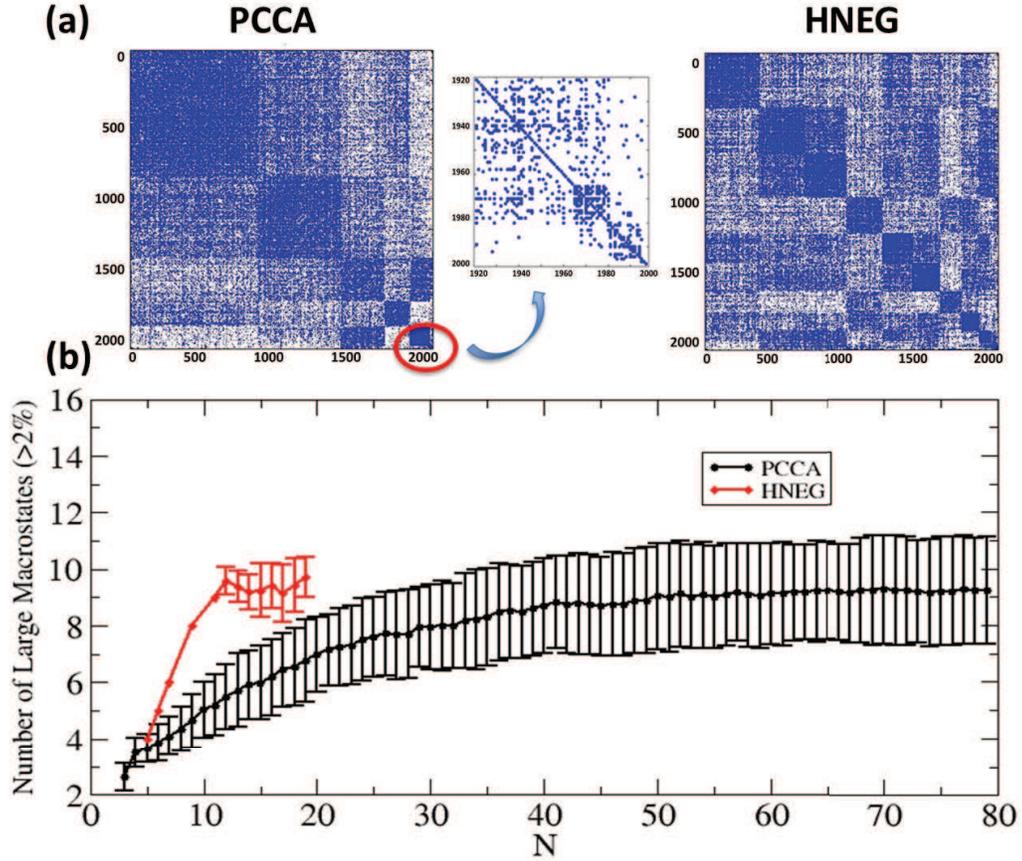}
\end{center}
\caption{
(a) An illustration of block diagonal structures of the microstate transition probability matrices.  There are 2000 microstates in total, and the matrices are permuted to group microstates that belong to the same macrostate together.  The results show that PCCA (left panel) tends to separate nearly disconnected small blocks first, while HNEG (right panel) focuses on identifying the well populated macrostates. The number of macrostates for both lumpings is $11$. (b) HNEG successfully identify the large macrostates (around $9$) with a much smaller total number  of macrostates ($<20$) compared to PCCA ($>80$).
}
\label{fig:trpzipc}
\end{figure}
\newpage

\begin{figure}[!h]
\begin{center}
\includegraphics[width=0.8\textwidth]{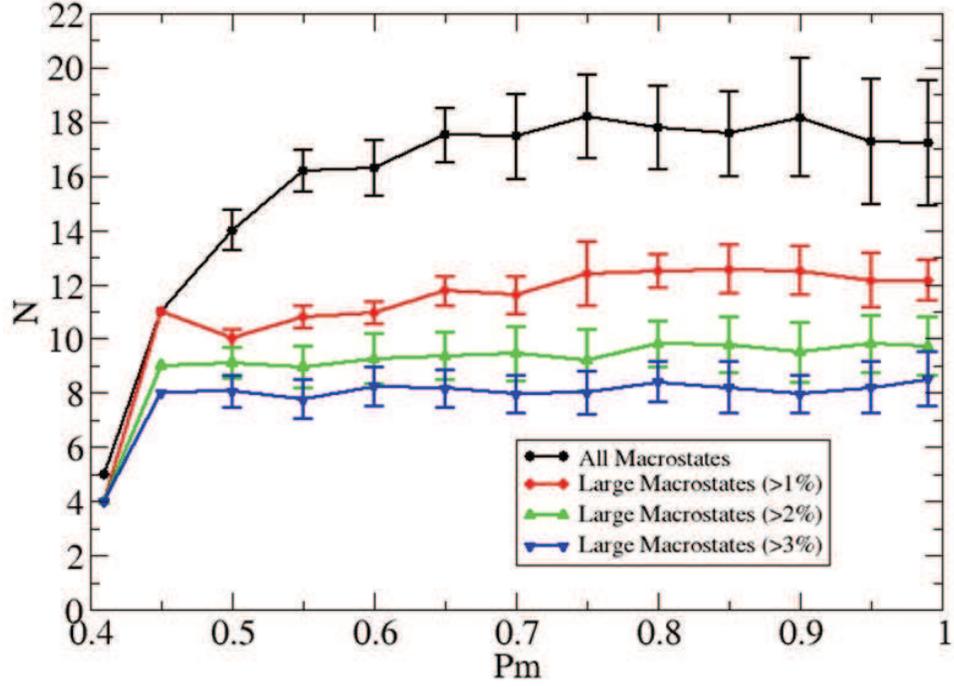}
\end{center}
\caption{
The Hierarchical Nystr\"{o}m method can robustly identify the large metastable macrostates with population greater than $1\%$(red), $2\%$(green) and $3\%$(blue), when varying the fraction of data that are included in the submatrix $A$ (see Sec II for details). The percentage of data we include($P_m$) in the submatrix $A$ is varied from $41\%$ to $99\%$. The number of large macrostates keeps the same after we include $50\%$ of the data or more.
}
\label{fig:trpzipd}
\end{figure}
\newpage

\begin{figure}[!h]
\begin{center}
\includegraphics[width=0.8\textwidth]{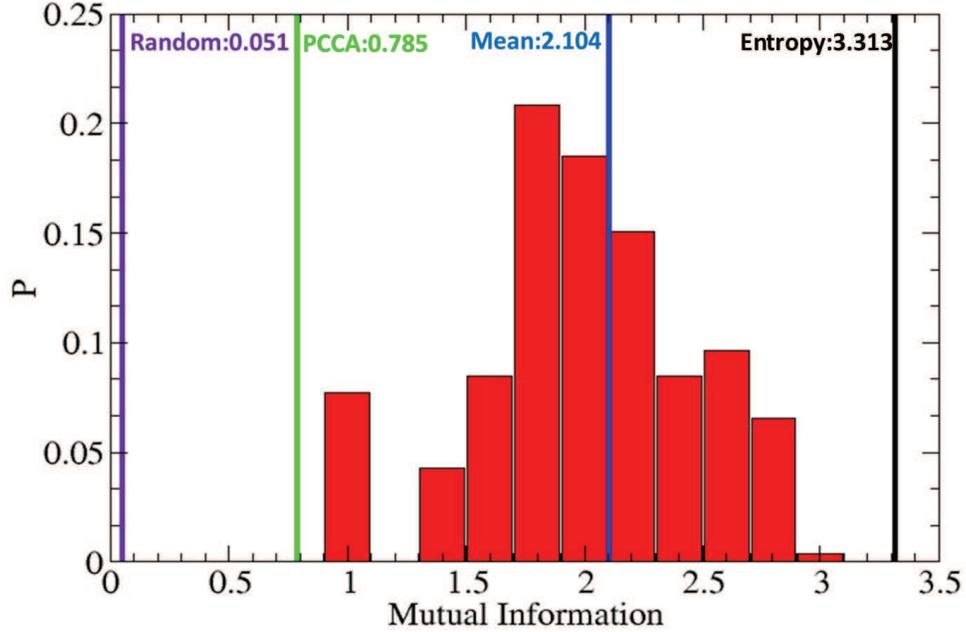}
\end{center}
\caption{
Histogram of pairwise mutual information (with a mean value of $2.104$) between the optimal lumping and all other lumpings (259 of them) obtained from the Nystr\"{o}m method by varying the level sets (red). For comparison, the mutual information between a lumping generated by PCCA (with 13 macrostates) and the optimal lumping is only $0.785$ (green). The entropy of the optimal lumping (upper limit of the mutual information) is $3.313$ (black), while the averaged mutual information between random lumpings (we produced 259 random lumpings) and the optimal lumping is $0.051$.}
\label{fig:trpzipf}
\end{figure}
\newpage

\begin{figure}[!h]
\begin{center}
\includegraphics[width=0.8\textwidth]{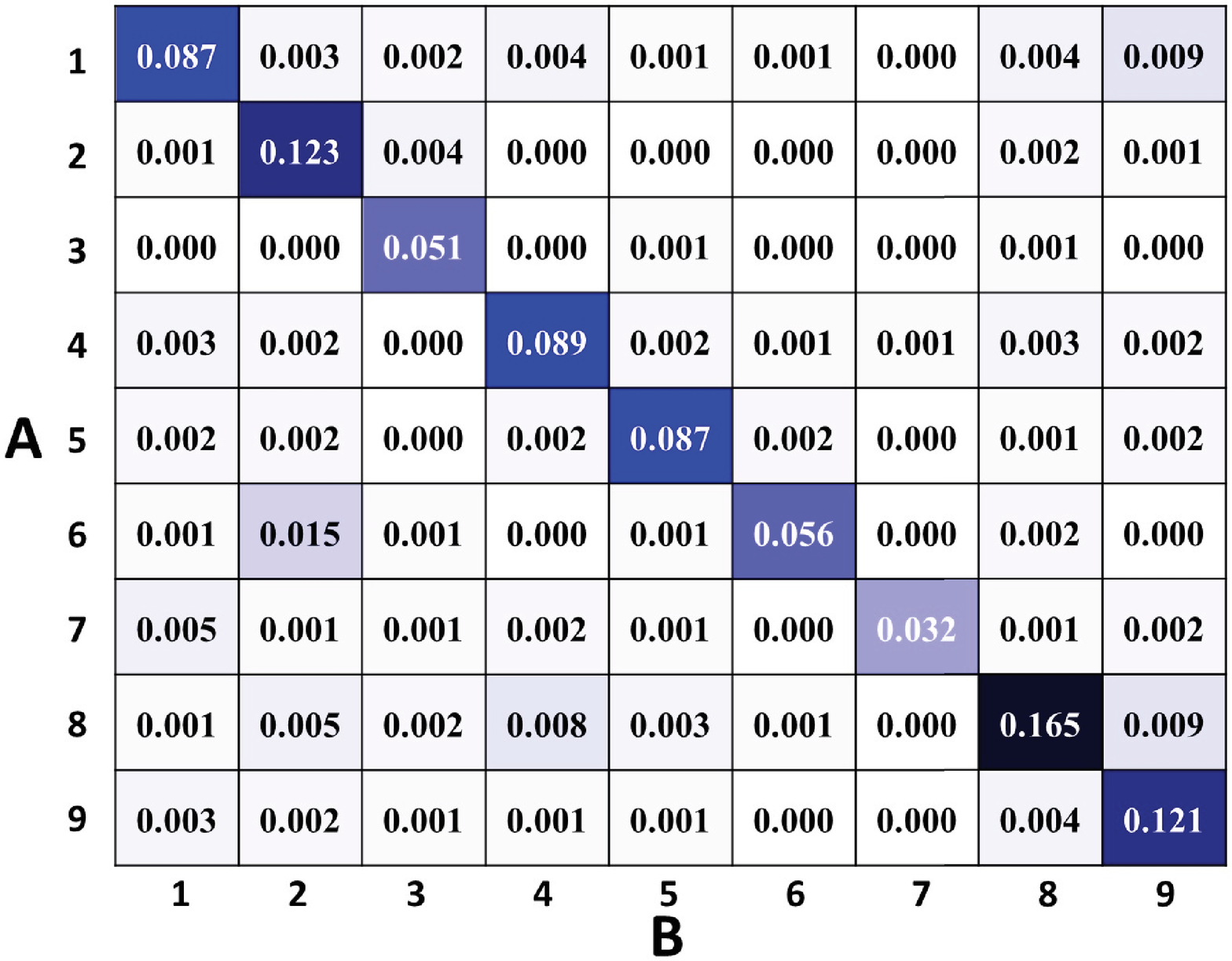}
\end{center}
\caption{
Illustration of the overlapping of microstates (or protein conformations) assigned to large macrostates in different lumpings. The optimal Lumping (A) is compared with a representative lumping (B) with mutual information at around $2.1$. Both of these lumpings contain 9 large macrostates states with population $>2\%$.  The joint probability matrix $P_{A, B}(i,j)$ indicates the overlapping of microstates assigned to macrostate $i$ in the optimal lumping A and macrostate $j$ in the representative lumping B. $P_{A, B}$ has large diagonal elements but small off-diagonal elements after permutation.  These results indicate that the large macrostates in the two lumpings share a relatively large fraction of identical microstates.
}
\label{fig:trpzipe}
\end{figure}
\newpage

\end{document}